\documentclass[%
longbibliography,
 reprint,
 amsmath,amssymb,
 aps,
 prl,
]{revtex4-1}

\usepackage{epstopdf}
\usepackage{graphicx}
\usepackage{dcolumn}
\usepackage{bm}
\usepackage{amsmath}
\usepackage{amssymb}
\usepackage{gensymb}
\usepackage{float}
\usepackage{svg}
\usepackage{hyperref}

\begin{document}

\preprint{APS/123-QED}

\title{Meniscus-Driven Modulation of Surface Wave Transmission Across Barriers}

\author{Zhengwu Wang}
\author{Guoqin Liu}
\author{Likun Zhang}%
 \email{zhang@olemiss.edu}
\affiliation{%
 National Center for Physical Acoustics and Department of Physics and Astronomy, University of Mississippi,\\
 University, Mississippi 38677, USA 
}%

\date{\today}

\begin{abstract}
Meniscus oscillations at interfaces between liquids, solids, and air significantly impact fluid dynamics and control. While idealized models exist, experimental data on capillary–gravity wave scattering involving meniscus effects remain limited. In this study, we systematically measured wave transmission past a surface-piercing barrier, focusing on meniscus and contact-line effects. By varying the barrier’s surface properties and the wave frequencies, we explored how meniscus deformation influences wave transmission. The results are compared with simulations and limiting-case theories. We find that the meniscus water column beneath the barrier enhances coupling and increases transmission, while surface bending suppresses it by constraining motion. These competing effects explain the observed frequency and contact angle dependencies. Our findings provide insights into how meniscus dynamics govern surface wave behavior.
\end{abstract}

\maketitle


\textit{Introduction} -- 
Capillary phenomena involving surface tension effects have long been a subject of interest in fluid physics and in normal-gravity and microgravity applications \cite{zhang2013capillary,marr2001passive,lowry2007fixed,bostwick2009capillary,bostwick2015stability,berhanu2020capillary}. The contact line and meniscus induced by surface tension are ubiquitous in various fluid structures, such as sessile drops  \cite{vukasinovic2007dynamics,fayzrakhmanova2009stick,sharp2012resonant}, bubbles  \cite{harazi2019acoustics}, and liquid bridges \cite{marr2001passive,morse1996capillary}. Their influence on the dynamics and stability of these structures has been extensively studied \cite{mei1973damping,hocking1987damping,henderson1994surface,jiang2004contact,nicolas2005effects,kidambi2011frequency,huang2020streaming,kim2020capillary}. Beyond standing surface waves \cite{perlin2000capillary}, the response of traveling capillary-gravity waves to contact line and meniscus effects also merits attention \cite{scott1978waves,benjamin1979gravity,cocciaro1991capillarity,perlin2000capillary,kidambi2009meniscus,shao2021role,monsalve2022space}. This study investigates a different fundamental problem: How does meniscus affect the transmission of traveling capillary-gravity waves through a surface-piercing barrier? This problem is relevant to everyday phenomena, such as ripples interacting with surface obstacles, as well as to applications in fluid control.

\begin{figure}[b]
    \includegraphics[width=\columnwidth]{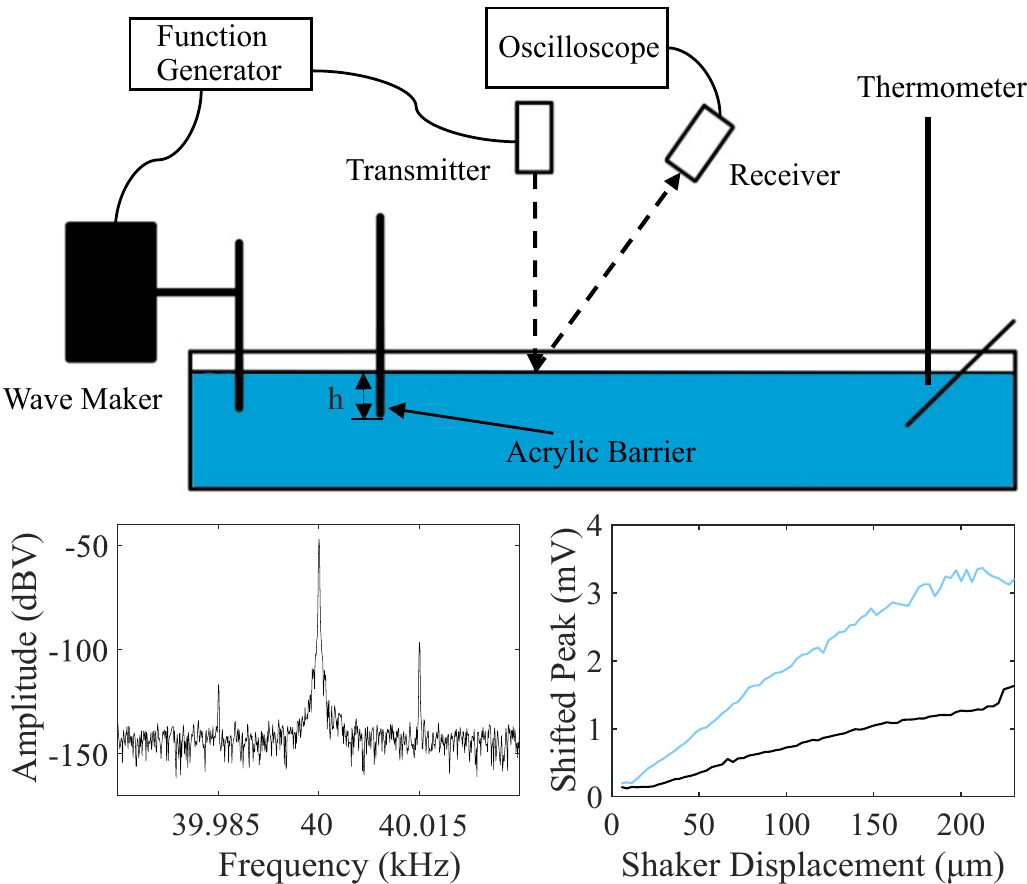}
    \caption{\label{fig1} Experimental setup: (a) Diagram of designed system using acoustic measurement to measure amplitude of capillary-gravity waves generated by a paddle wavemaker. (b) Typical spectrum of 40-kHz acoustic signal modulated by surface waves of 15 Hz to have two shifted peaks. (c) Typical dependence of a shifted peak amplitude on the displacement of paddle with (black) and without (blue) the barrier to obtain transmission coefficient $|T|$ by a ratio between the two.}
\end{figure}

The fundamental problem of wave transmission and reflection at a barrier was analytically studied under gravity wave conditions as early as the 1940s by Dean and Ursell \cite{dean1945reflexion,ursell1947effect}. In 1968, Evans \cite{evans1968influence} introduced surface tension effects, enabling consideration of contact line effects. Subsequently, in the 1980s, Rhodes-Robinson and Hocking \cite{rhodes1982note,hocking1987reflection,rhodes1996effect} developed theoretical models for the transmission and reflection of capillary-gravity waves. Zhang and Thiessen \cite{zhang2013capillary} analytically examined the transmission and dissipation of capillary waves through a ring-constrained liquid cylinder in zero gravity. However, despite the theoretical advances, experimental data to compare with these models remain sparse, particularly because theoretical models have assumed idealized scenarios, such as the absence of meniscus effects. Obtaining experimental data to address these gaps is essential.

In this study, we use acoustic measurements \cite{michel2016acoustic} to investigate the transmission coefficient of capillary-gravity waves through a surface-piercing barrier, explicitly accounting for the meniscus effect. All measurements are conducted under fixed contact line conditions, avoiding the complexities associated with mobile contact lines \cite{dussan1979spreading,cocciaro1993experimental,ting1995boundary}. We examine the influence of the meniscus by varying both the contact angle and the frequency of the surface waves. By altering the height and surface properties of the barrier, we systematically manipulate the meniscus to study its effects. Our experimental approach establishes controlled conditions and provides results suitable for comparison with theoretical and numerical models under idealized assumptions. We provide interpretations to elucidate the underlying physics and the dependencies on relevant parameters.

\textit{Experimental methods} -- The experimental setup is illustrated in Fig.~\ref{fig1}(a). The inner dimensions of the experimental tank are $106 \times 6.8 \times 11~\text{cm}^3$ (L $\times$ W $\times$ H), and it was filled with distilled water to a height of $9.2~\text{cm}$. The barrier used for transmission measurements was a $6.8~\text{cm}$ wide, finite-thickness acrylic sheet mounted on a height-adjustable translation stage capable of a maximum resolution of $0.01~\text{mm}$. A paddle wavemaker, driven by a Frederiksen model 2185 shaker, generated surface waves of specific frequencies and was positioned approximately $17~\text{cm}$ from the barrier, over a distance of more than ten wavelengths for a $15~\text{Hz}$ surface wave (a typical frequency range of $5-25~\text{Hz}$ in our measurements of capillary-gravity waves). The distance between the barrier to transmitter has also designed to be $17~\text{cm}$. The wave amplitude is much smaller than the wavelength for linear wave approximation (see SI Section 4.1).

The transmitter and receiver setup was adapted from an acoustic method developed in \cite{michel2016acoustic} to measure the amplitude of capillary-gravity waves. In this setup, ultrasound reflects off the liquid surface, producing a spectrum of the received signal that shows two frequency-shifted peaks [Fig.\ref{fig1}(b)] around the incident source frequency ($f_0=40$~kHz), plus or minus the surface wave frequency $f_w$. These shifted peaks at $f_0 \pm f_w$ result from modulation of the sound signal when reflecting off the surface waves. The fast Fourier transform (FFT) of each signal was taken, and the transmission coefficient was determined by calculating the ratio of the amplitude of the modulated signal's shifted component at $f_0+f_w$ [Fig.~\ref{fig1}(c)] between cases with ($A_w$) and without ($A_{w/o}$) the barrier immersed in the water, namely, $|T| = A_w / A_{w/o}$. Each signal was recorded over a 12-second period, containing 2 million data points. The measurement was repeated five times, with a 5-minute waiting period between measurements for the system to reach a steady state. The five repeated measurements were used to average the transmission coefficients and estimate the associated uncertainty (see SI Section 1.2).

One constraint on the validity of this measurement technique is that the displacement amplitude of the water waves, $\eta$, must be much smaller than the acoustic wavelength ($k_a \eta \ll 1$, where $k_a$ is the wavenumber of the acoustic wave). Maintaining a linear response in the measurement was therefore essential. An accelerometer installed near the base of the paddle wavemaker was used to monitor the amplitude of the wavemaker based on its acceleration and frequency. Validation tests [Fig.~\ref{fig1}(c)] showed that when the wavemaker amplitude exceeded around $150~\mu\text{m}$, significant errors and variations began to appear in the shifted peak response. Based on these findings, all subsequent measurements in this study were performed with a wavemaker amplitude of no more than 120~µm, leading to an estimated surface wave amplitude ranging from 13~µm at 25~Hz to 300~µm at 5~Hz (see SI section 4.1).

\begin{figure*}
    \centering
     \includegraphics[width=0.95\textwidth]{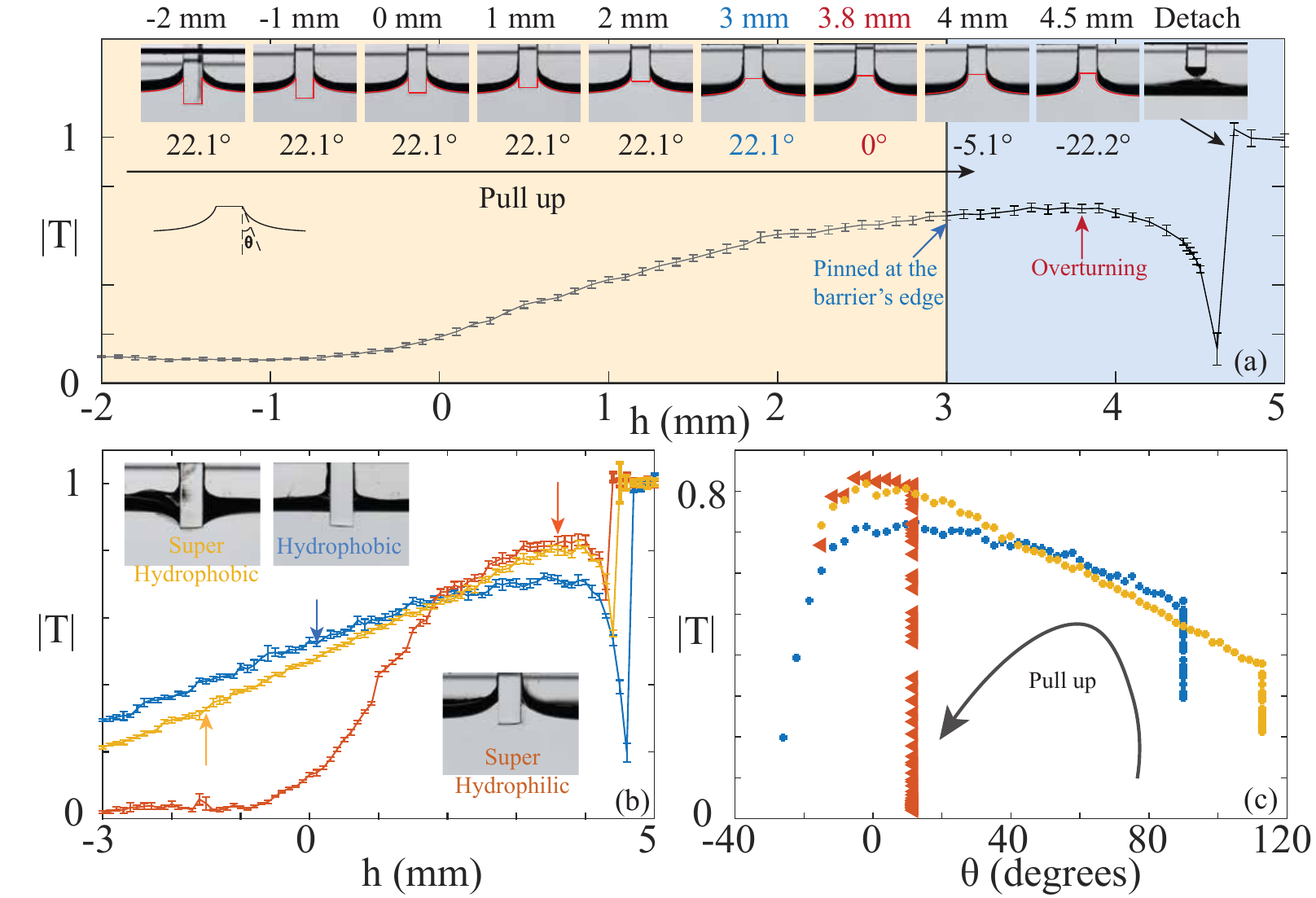} \\
 \caption{\label{fig2} 
 Experimental measurements of the transmission coefficient $|T|$ by varying the barrier height $h$ from $-2.0~\text{mm}$ to $5.0~\text{mm}$ in 0.1 mm increments under a 15 Hz capillary-gravity wave setup. The height $h$ is measured relative to the water level far from the meniscus [Fig.~\ref{fig1}(a)]. (a) Dependence of $|T|$ on $h$ as the barrier is pulled up from immersion (yellow background) to pinning the contact line at the barrier edge (blue background), until detachment from the liquid. Insets show side-view images of the static meniscus at the barrier, along with theoretical meniscus profiles (red) from Eq.~(\ref{eq:eq1}). (b) Dependence of $|T|$ on $h$ for three barriers with surfaces coated to be superhydrophobic, hydrophobic, and superhydrophilic, respectively (see insets for $h = -2$ mm). All measurements were conducted by pulling up the barrier, and the arrows indicate the height at which the contact lines begin to pin at the barrier's edges. (c) Dependence of $|T|$ on contact angle $\theta$ using the data from (b), where the black arrow indicates the contact angle trend as the barrier is pulled up from $-3.0~\text{mm}$ to $5.0~\text{mm}$.}
\end{figure*}

\textit{Observations of the meniscus effect } -- 
The typical measurement we constructed is by pulling the barrier up from under the water surface [see Video S1] to observe the transmission variation with the barrier height $h$ (where $h=0$ is defined as the bottom of barrier at the level of water away from the meniscus) for a 15 Hz surface wave. The transmission coefficient initially increased with the barrier height when raising the barrier from $-2.0~\text{mm}$ up to $3.0~\text{mm}$, a height at which the contact lines begin to pin at the edges of the barrier [Fig.~\ref{fig2}(a)]. 

With the contact line pinned at the edges of the barrier at $h=3.0~\text{mm}$, we further raised the barrier height until the barrier is detached from the water surface at 4.7 mm. During this process, the transmission initially increased, then saturated, and finally decreased. In particular, the transmission drops significantly at the very large height.  Since the contact lines are pinned at the barrier edge, no portion of the barrier remains submerged. The observed changes of transmission results from the changes of the meniscus, a phenomenon that has not been investigated by any prior studies.

We further measured the transmission dependence on the height for barriers with a hydrophobic or hydrophilic coating. Three different coatings were applied for a hydrophobic barrier with a zero meniscus, a superhydrophobic barrier with a negative meniscus, and a superhydrophilic barrier with a positive meniscus [insets of Fig.~\ref{fig2}(b)]. Following from the measurements by pulling up the barrier, the transmission were measured as a function of the barrier height for these three barriers. By comparing all the three types of barrier, we observed the same phenomenon as the uncoated barrier in Fig.~\ref{fig2}(a) that further raising the barrier height after the contact lines pinned at the edges (indicated by the arrows in Fig.~\ref{fig2}(b)) would continue increasing the transmission until a certain height where the reduction of the transmission begins to appear [Fig.~\ref{fig2}(b)]. 

The transmission in Fig.~\ref{fig2}(b) differs between the three cases due to the variation in the heights at which the fluid begins to pin at the barrier's edge, as well as the secondary difference in the coating at the bottom of the barrier. In the superhydrophobic case, the fluid begins to pin at the edge of the barrier at a negative height $h$. In this case, the measurements show that increasing the barrier height from this negative value enhances transmission until the previously described suppression effect begins to take place.

\textit{Interpretations} -- We now interpret the measured transmission. As demonstrated in our experiments (see SI Section~4.2),  the variation in contact angle remains within 3.5\(^{\circ}\), which is significantly smaller than the contact angle hysteresis range of approximately 50\(^{\circ}\). As such, the contact lines remain pinned at the side walls of the barrier throughout, resulting in no measurable dissipation. The initial reduction in transmission observed in Fig.~\ref{fig2}(a) thus indicates enhanced reflection: a portion of the wave energy is transmitted, while the remainder is reflected. The significant suppression of transmission at large negative immersion depths arises from the fact that the immersed barrier interferes with the oscillatory flow, which is predominantly located in the upper layer of the fluid for surface waves. This disruption prevents efficient coupling across the barrier, leading to strong reflection and reduced transmission at a large negative height $h$ (yellow regime in Fig.~\ref{fig2}(a)). During the process of contact lines pinned at the side walls of the barrier, the meniscus maintains its shape and static contact angle. 

Our focus is to interpret how the meniscus affects transmission when the contact lines are pinned at the edge of the barrier, causing it to first increase and then decrease with the rise of the meniscus (blue regime in Fig.~\ref{fig2}(a)). On one hand, the presence of the extra water column above the background water level of no meniscus should enhance the coupling of surface oscillations from the incident side to the transmitted side, as the energy of the oscillation is mainly localized in the top layer of the fluid. On the other hand, the surface bending due to the inclination of the meniscus slope leads to more constrained surface motion, which should enhance reflection and suppress transmission. The interplay between these two factors explains the non-monotonic behavior: As the meniscus rises from a small slope to a large slope, the water column begins to form and increases transmission. However, beyond a certain point, the slope of the meniscus becomes sufficiently steep that bending-induced suppression takes over, leading to a decrease in transmission.

With this interpretation, for a negative meniscus, where the meniscus bends downwards below the background water level (i.e., the super-hydrophobic case in Fig~.\ref{fig2}(b)), the bending of the meniscus should similarly reduce the transmission. The water column formed by a positive meniscus does not appear in the case of a negative meniscus. Instead, the negative meniscus can suppress the local water level beneath the barrier. This reduces the effective cross-sectional area available for oscillatory flow, potentially further weakening the coupling between the incident and transmitted sides. As a result, for contact lines pinned at the edge of the barrier, when raising the barrier from a negative height to a zero height by releasing the water from a suppressed state to a flat state, we would expect the transmission to increase. This is exactly what we observed in Fig.~\ref{fig2}(b).

\textit{Contact angle dependence and interpretations} -- 
We now explore factors and parameters that influence the transmission, particularly those related to the meniscus effects. First, we matched the meniscus profile from our experimental images with the meniscus profile derived from pressure balance [insets of Fig.~\ref{fig2}(a)]. The meniscus profile resulting from the pressure balance is described by the horizontal coordinate \( x \) as a function of the vertical coordinate \( z \) as \cite{gennes2004capillarity}:
\begin{equation}
    x - x_0 = a \cosh^{-1} \left( \frac{2a}{z} \right) - 2 a \sqrt{1 - \frac{z^2}{4 a^2}}
    \label{eq:eq1}
\end{equation}
where \( a = \sqrt{\sigma / \rho g} \) is the capillary length (2.7 mm in our experiments), \( \rho = 1000 \, \text{kg/m}^3 \) is the density of water, \( g = 9.81 \, \text{m/s}^2 \) is the gravitational acceleration, and \( \sigma = 72 \, \text{dyne/cm} \) is the surface tension at the water–air interface. The meniscus profiles extracted from the experimental images were fitted using Eq.~(\ref{eq:eq1}) through a nonlinear fitting method, with \( x_0 \) and \( a \) treated as free parameters (see SI Section~2.1). These fitted profiles were then used to determine the capillary rise at the barrier surface (denoted by \( h_c \)) and the contact angle \( \theta \), as
\begin{equation}
    \sin \theta = 1 - \left( \frac{h_c}{\sqrt{2}a} \right)^2.
    \label{eq:eq2}
\end{equation}

In the case of contact lines pinned at the edges of the barrier, we measured the contact angle with respect to the vertical extension of the barrier [insets of Fig.~\ref{fig2}(a)] so that the contact angle are still described by Eq.~(\ref{eq:eq2}), where the capillary rise equals to the barrier height $h$. Raising the meniscus reduces the contact angle. At $h$=3.8~mm, the contact angle reaches $0^{\circ}$ and the surface normal turns to horizontal. Further raising the barrier leads to a negative contact angles for the meniscus profiles becoming an overturned shape and the two meniscus profiles form an ``I-shaped'' water column underneath the barrier [inset of Fig.~\ref{fig2}(a)]. 

A significant suppression of transmission occurs at very negative contact angles [Fig.~\ref{fig2}(a)]. We attribute this to the overturning of the meniscus interface: as the contact angle passes through zero and the interface flips, the incident surface wave propagating along the water column is refracted in the opposite direction relative to its initial trajectory. This reversal in wave propagation, induced by the steep meniscus inclination, restricts the forward propagation of the incident waves, leading to increased reflection and a corresponding decrease in transmission. Our measurements in Figs.~\ref{fig2}(a) and (b) demonstrate that this suppression of transmission, caused by the overturned meniscus at very negative contact angles (large $h$), is comparable in magnitude to the suppression observed at large negative $h$ due to the immersion effect.

The meniscus is described by the capillary rise $h_c$ and the capillary length $a$. The ratio between the two determines the contact angle $\theta$ as in Eq.~(\ref{eq:eq2}). As such, we also present the transmission dependence on the contact angle (Fig.~\ref{fig2}(c)), which displays what we observed in Fig.~\ref{fig2}(b) but in the perspective of contact angle: As the barrier is raised, the fluid maintains a fixed contact angle while the contact lines are pinned to the side wall of the barrier, and transmission continues to increase due to reduced barrier immersion. Once the contact line becomes pinned at the edge of the barrier, further raising the barrier leads to a decrease in the contact angle, causing the transmission to begin to decrease. This behavior occurs whether the meniscus starts in a negative state (\(\theta > 90^\circ\)) or a positive state (\(\theta < 90^\circ\)). Eventually, as the contact angle approaches \( 0^\circ \) and the meniscus transitions to an overturned state, the transmission begins to saturate and then decrease, as previously discussed. This behavior is consistent across all coating types.

\begin{figure*}
     \centering        
        \includegraphics[width=0.98\textwidth]{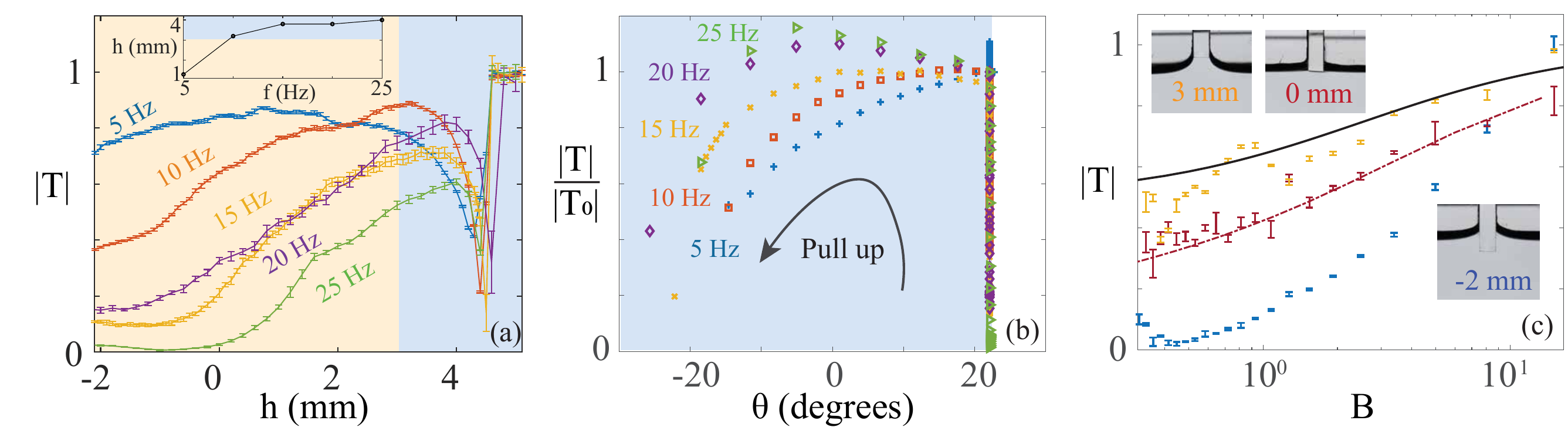}         
        \caption{\label{fig3} Dependence of transmission $|T|$ on wave frequency, compared with numerical simulations and theoretical predictions. 
        (a) Transmission as a function of barrier height for 5 frequencies, with the yellow-blue boundary  at \( h = 3 \, \text{mm} \) indicating the transition when contact lines begin to pin at the barrier edges; the inset shows the height of maximum transmission for each frequency. 
        (b) Transmission vs. contact angle for the data in (a), normalized by the transmission coefficient $|T_0|$ at \( h = 3 \, \text{mm} \). 
        (c) Transmission vs. Bond number $\text{B}$ for frequencies from 5 to 25 Hz, comparing measurements (immersed barriers at \( h = -2 \, \text{mm} \) and non-immersed barriers at \( 0 \, \text{mm} \) and \(3 \, \text{mm} \)), simulations (0 mm, red dotted curve), and theoretical results (infinitesimal barrier, no meniscus, black curve).
        }
\end{figure*}

\textit{Frequency influence and interpretations} -- 
While the meniscus shape is fixed for a given contact angle determined by the ratio of the capillary rise \( h_c \) to the capillary length \( a \) (see Eq.~(\ref{eq:eq2})), the surface wave wavelength varies with wave frequency. In principle, one could introduce dimensionless parameters such as \( h_c/\lambda \) or \( a/\lambda \) to characterize the relative scale between the meniscus and the wavelength \( \lambda \). As such, frequency becomes an additional important parameter influencing the meniscus effect on wave transmission. All measurements conducted so far were for a 15 Hz wave. We have now extended the measurements to cover a frequency range of 5-25 Hz to investigate how frequency affects transmission behavior. By raising the uncoated barrier, the measured transmission [Fig.~\ref{fig3}(a)] exhibits a consistent trend across all frequencies: transmission first increases and then decreases as the barrier height increases, up to the point of fluid detachment. However, the increase-decrease transition occurs at different heights for different frequencies, with the transition occurring at a greater barrier height for higher frequencies [inset of Fig.~\ref{fig3}(a)].

We interpret the increase-decrease transition dependence on frequency through the relative contributions of the two competing mechanisms: the enhancement due to elevated water column coupling and the suppression from meniscus bending. At lower frequencies, the waves experience a greater change in meniscus curvature over a larger wavelength scale, leading to an amplification of the meniscus bending effect, which suppresses transmission even with a smaller capillary rise. In contrast, higher-frequency waves, dominated by surface tension, are more sensitive to water column coupling as a capillary phenomenon, which in turn enhances transmission even at a larger capillary rise. This explains the observed increase-decrease transition occurring at a greater height for higher frequencies.

In Fig.~\ref{fig3}(a), the fluid consistently begins to pin at the barrier edges at the same height of 3 mm, regardless of frequency. For submerged barrier regime (yellow background, $h<3$~mm), we observed that increasing the frequency generally result in reduced transmissions, as the barrier size effects are enhanced at higher frequency waves of shorter wavelengths, which can be described by the dimensionless parameters $h/\lambda$ with the $h$ dominant by the submerged depth and $w/\lambda$ with $w$ being the width of the barrier. However, for contact lines pinned at the barrier edges (blue background, $h>3$~mm), the interplay between the meniscus and the frequency leads to no clear trend for the dependence of the transmission on frequency in this scenario. 

Considering this complex regime, we normalize the measured transmission \( |T| \) at each frequency by their respective values at the transition when the contact lines begin to pin at the edges of the barrier at the height of 3 mm (yellow-blue boundary in Fig.~\ref{fig3}(a)), denoted by \( |T_0| \). This normalization eliminates the transmission difference when the contact line just begins to pin at the barrier edge. We then plot the normalized transmission \( |T|/|T_0| \) as a function of contact angle in Fig.~\ref{fig3}(b), which reflects the effect of the meniscus on the trend of the normalized transmission when varying the contact angle by raising the meniscus under different frequencies. We observe a trend that increasing the frequency raises the normalized transmission while decreasing the frequency suppresses the normalized transmission. This tread of normalized transmission with frequency is exactly what we would expect based on the interpretation that the transmission is enhanced by water column coupling that dominates at high frequencies and is suppressed by the meniscus bending that dominates at low frequencies. The critical contact angles for the transition between increasing and decreasing transmission with respect to the contact angle correspond to the critical $h$ values shown in inset of Fig.~\ref{fig3}(a). 

Lastly, changes in frequency also alter the interplay between surface tension and gravity, which can be characterized by the dimensionless Bond number:
\begin{equation}
\text{B} = \frac{\rho g}{\sigma k^2}.
\end{equation}
where $k$ is the wavenumber of the surface wave. A lower B corresponds to higher frequency with the dispersion relation $\omega^2 = gk +\sigma k^3 / \rho$ in the deep water approximation. To further characterize the dependence on the Bond number, we measured the transmission by systematically varying the frequency from 5 Hz to 25 Hz at fixed heights [Fig.~\ref{fig3}(c)]. For three different heights, the measurements show that the transmission generally decreased with increasing frequency, consistent with enhanced surface tension effects at higher frequencies and conveying the description of the dependence by the Bond number [see Video S3].

For comparison, we conducted complementary fluid dynamics simulations for linear waves on a flat meniscus in potential flow under the inviscid, irrotational, and incompressible assumptions in which pinned contact line conditions were applied (see SI section 5.1). Our simulated results for the transmission dependence on the Bond number [Fig.~\ref{fig3}(c)] agree well with our measurements under the same setting of a zero meniscus, reinforcing our experimental results and the pinned contact line conditions in our measurements. We also compared our results with a theoretical prediction of the transmission through an infinitesimal barrier where the meniscus is flat \cite{zhang2013capillary,liu2025cylinder} (see SI section 5.2). The theoretical prediction and our experimental measurements reinforce each other on the general increase of the transmission with the Bond number [Fig.~\ref{fig3}(c)], though our measurements with a finite-width barrier are in general smaller than the theoretical prediction with a zero-width barrier. 

\begin{figure}
     \centering    
     \includegraphics[width=0.48\textwidth]{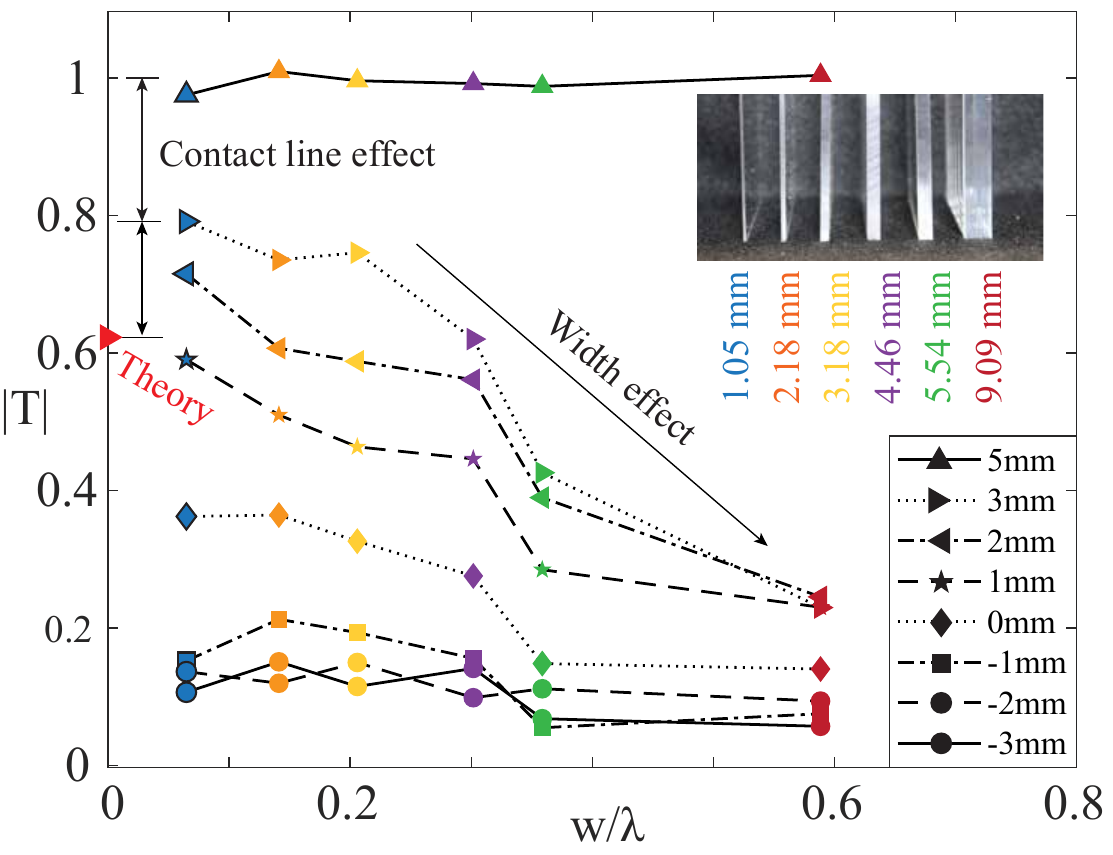} 
        \caption{\label{fig4} Transmission variation with barrier width $w$ for uncoated barriers of 6 different widths (insets). The width $w$ is normalized by wavelength $\lambda$= 15.4 mm of the measured 15~Hz wave. The measurements were for both immersed (ranging from -3 mm to 2 mm) and non-immersed (3 mm) barriers. The red diamond dot indicates a theoretical prediction of the transmission $|T|=$ 0.62 for an infinitesimal barrier and no meniscus.}
\end{figure}

\textit{Remarks on the barrier size and pinned contact line effects} –
Our measurements were limited to barriers with a fixed width of $w$ = 3.18 mm. We now extend measurements for six uncoated barriers with different widths ranging from 1 mm to 9 mm, all taken at a single frequency of 15 Hz. The width range spans from 0.06 to 0.6 times the wavelength $\lambda = 15.4$ mm of the wave. The measured transmission for these six barriers, as a function of the barrier width normalized by the wavelength, is shown in Fig.~\ref{fig4} for both immersed and non-immersed barriers beneath the liquid surface. The measured transmission generally increases with a reduction in barrier width and immersed depth for the measured frequency, due to the reduced finite-size effect.

Our focus in Fig.\ref{fig4} is that when the width of the barrier is much smaller than the wavelength and the barrier has no submerged portion—effectively approaching the limit of an infinitesimal barrier—the transmission is reduced from unity to a smaller value of 0.79. This reduction experimentally confirms a prediction of transmission reduction due to pinned contact lines \cite{zhang2013capillary,liu2025cylinder} (see SI section 4), which depends on the Bond number. Our measured transmission value of 0.79 [Fig.~\ref{fig4}] is higher than the theoretically predicted value of 0.62 (with \text{B} = 0.81), because the experiment involves a positive meniscus, whereas the theoretical model assumes a flat surface \cite{liu2025cylinder} .

Lastly, we remark on the complex multiple-parameter dependencies due to the combination of meniscus effect and barrier's finite-size effect. For wave transmission in general, one might simply expect smaller transmission with a larger barrier. The finite-size effect depends solely on the relative scale between the barrier size and the wavelength, which is \( w/\lambda \) and \( h/\lambda \) herein. Nevertheless, due to the additional meniscus effects, even when \( w/\lambda \) and \( h/\lambda \) are held fixed, variations in frequency still affect other frequency-dependent parameters that influence the meniscus effect, such as the Bond number and the relative scale between the meniscus and the wavelength, as identified in Fig.~\ref{fig3}. Additionally, the width of the barrier affects the width of the meniscus-induced water column beneath the barrier, which can influence water column coupling—an effect that warrants further investigation.

\textit{Summary} -- 
We have systematically investigated the influences of the meniscus effect on the surface wave transmission coefficient by varying the meniscus shape, the wave frequency, and the geometric size of barriers. Our measurements provide the first experimental data addressing this problem under realistic conditions. Numerical simulations and existing theoretical predictions are compared with our experimental results in limiting cases.

A particularly novel observation was the enhancement of transmission when raising the small-slope meniscus and the suppression of transmission when further raising the large-slope meniscus before detachment, a phenomenon not predicted by any existing theoretical models. We interpret this as follows: the elevated water column formed beneath the barrier enhances transmission, whereas the increased inclination of the meniscus suppresses it. The influence of the former mechanism becomes more pronounced at higher frequencies, while the latter dominates at lower frequencies. The interplay between these two effects accounts for the observed trend in transmission as a function of contact angles and excitation frequency.

We revealed factors that influence the meniscus effects: (a) the meniscus shape itself, described by the capillary rise normalized by the capillary length, \( h_c/a \), which alternatively determines the contact angle \( \theta \); (b) the relative effect between the surface wave scale and the meniscus extent, characterized by \( h_c/\lambda \) or \( a/\lambda \); (c) the relative effect between surface tension and gravity, described by the Bond number \( \text{B} \). Additional parameters are the geometric dimensions of the barrier, such as its width and immersion depth, which can be normalized by the surface wavelength as \( w/\lambda \) and \( h/\lambda \), respectively. 

The combined effects of meniscus shape, frequency, and the geometry introduce multiple parameter dependencies that complicate the problem, requiring a comprehensive theoretical or numerical framework. 
	
	\begin{acknowledgments}
		This work is supported by NSF Fluid Dynamics Program, Grant No. 2306106. We acknowledge preliminary work conducted by Robert Lirette and Zheguang Zou through the support of a seed grant from MS NASA EPSCoR.
	\end{acknowledgments}

\nocite{*}

\providecommand{\noopsort}[1]{}\providecommand{\singleletter}[1]{#1}%

\end{document}